\documentclass{tex-library/jfm}

\usepackage{graphicx, amsmath, amssymb, amsfonts, mathtools, mathrsfs, color}
\usepackage{natbib, comment, todonotes}
\usepackage{xcolor}
\usepackage{soul}
\usepackage[colorlinks]{hyperref}
\usepackage[capitalise]{cleveref}
\usepackage{bm}
\usepackage[english]{babel}
\usepackage[toc,page]{appendix}
\usepackage{url}
\usepackage[normalem]{ulem}
\usepackage{mathbbol}
\usepackage{upgreek}
\usepackage{bbm}

\newcommand{\pd}[2]{ \frac{ \partial #1}{ \partial #2 } }
\newcommand{\ppd}[2]{ \frac{ \partial^2 #1}{ {\partial #2}^2 } }

\newcommand{\Lap}{\grad^2}

\newcommand{\bvec}[1]{\ensuremath{\boldsymbol{#1}}}
\newcommand{\grad}{\nabla}

\newcommand{\mac}[1]{{#1}}
\newcommand{\nick}[1]{{#1}}

\newcommand{\uu}{\bvec{u}}

\newcommand{\T}{\theta}
\newcommand{\Pra}{\text{Pr}}
\newcommand{\Ra}{\text{Ra}}

\newcommand{\Nu}{\text{Nu}}

\newcommand{\second}{$2^\textnormal{nd}$}

\newcommand{\updel}{\Updelta}

\newcommand{\Cr}{\mbox{\textit{Cr}}}
\newcommand{\surf}{U}
\newcommand{\indp}{\mathbbm{1}_P}

\title{Covering convection with a thermal blanket: numerical simulation and stochastic modeling}

\author{Jinzi Mac Huang\aff{1,2}\corresp{\email{machuang@nyu.edu}}}

\affiliation{
\aff{1}NYU-ECNU Institute of Physics and Institute of Mathematical Sciences, New York University Shanghai, Shanghai, 200124, China
\aff{2}Applied Math Lab, Courant Institute, New York University, New York, NY 10012, USA
}

\begin{document}

\maketitle

\begin{abstract}
Adding moving boundaries to convective fluids is known to result in nontrivial and surprising dynamics, \mac{leading to spectacular geoformations ranging from the kilometer-scale karst terrains to the planetary-scale plate tectonics.}  \mac{On one hand, the moving solid alters the surrounding flow field, but on the other hand, the flow modifies the motion and shape of the solid. This leads to a two-way coupling that is significant in the study of fluid-structure interactions and in the understanding of geomorphologies.} In this work, we investigate the coupling between a floating plate and the convective fluid below it. Through numerical experiments, we show the motion of this plate is driven by the flow beneath. However the flow structure is also modified by the presence of this plate, leading to the ``thermal blanket" effect \mac{ where the trapped heat beneath the plate results in buoyant and upwelling flows that in turn push the plate away}. By analyzing this two-way coupling between moving boundary and fluid, we are able to capture the dynamical behaviors of this plate through a low-dimensional stochastic model. Geophysically, the thermal blanket effect is believed to drive the continental drift, therefore understanding this mechanism has significance beyond fluid dynamics. 
\end{abstract}

\begin{keywords}
Authors should not enter keywords on the manuscript, as these must be chosen by the author during the online submission process and will then be added during the typesetting process (see http://journals.cambridge.org/data/\linebreak[3]relatedlink/jfm-\linebreak[3]keywords.pdf for the full list)
\end{keywords}

\section{Introduction}
The interior of Earth has fascinated generations of scientists \citep{plummer2001physical}. Among them, Leonardo da Vinci (1452-1519) was one of the pioneers who noticed the incessant geological movements of our planet, as he observed the presence of marine fossils in the mountains. We now know the continents of Earth do not stay in place and instead undergo tectonic motions, and thermal convection in Earth’s mantle is believed to be the driving force of these motions \citep{kious1996dynamic}. 

Thermal convection occurs when uneven temperatures of fluid lead to uneven density and buoyancy, so warm fluid rises while cold fluid sinks. The definition of fluids here can be very broad, as modern geologists confirm that even the mantle flows like fluids at a large time scale \citep{turcotte2002geodynamics}. The Prandtl number ($\Pra{}$) there, defined as the ratio between the mantle's kinematic viscosity and thermal diffusivity, is estimated to be around $10^{23}$ \citep{meyers1987encyclopedia}. 

The core of Earth is much warmer than its surface, and the destabilizing buoyancy is strong enough to drive mantle convection. As a measure of relative strength between buoyancy and viscous effects, the Rayleigh number ($\Ra{}$) is around $10^6$ in the mantle \citep{selley2005encyclopedia}. In the well-studied case of Rayleigh-B\'enard convection, such a high $\Ra{}$ is known to lead to turbulent fluid motions \citep{Ahlers2009}. With the mantle convecting like a fluid, its surface flow transports the continental plates resulting in their tectonic motions.

Due to the large spatial scale of Earth and the long time scale of mantle convection, the geophysical study of plate tectonics focuses on the current state of continents as well as predicting its important consequences like earthquakes \citep{plummer2001physical}. On the other hand, numerical simulations \citep{howard1970self,Whitehead1972,Whitehead2015,mao2019dynamics,mao2021insulating,Whitehead2022} and lab-scale experiments \citep{elder1967convective,zhang2000periodic,Zhong2005,zhong2007a,zhong2007b,whitehead2011cellular} have proven to be an effective means of understanding the dynamics of plate tectonics. 

\mac{Early experiments of \cite{elder1967convective} showcase how one can recover the tectonic motion in the lab, where a paraffin fluid layer was used to model the mantle while a thin sheet of plastic floating on top served as a model continental plate. When the paraffin is heated from below, convection occurs and the plate moves due to shearing of the convective flow beneath. Such a simple experimental setup also displays nontrivial dynamics, as \cite{zhang2000periodic} observed the plate moves periodically between the two bounding walls of the fluid surface. Through more detailed investigations \citep{Zhong2005,zhong2007a,zhong2007b}, the size of floating plate is shown to strongly affect the plate motion, where small plates display periodic motion while large plates stay trapped in the middle of the fluid surface. Placing a moving heat source on top of a thermally convecting fluid also yields plate motions, and this is experimentally investigated by \cite{howard1970self,Whitehead1972}. Although the geometry, physical parameters, and time-scales presented in these works are very different from the mantle convection, they reveal surprising dynamics and most importantly, provide invaluable insight into the fluid-structure interaction mechanism behind continental drift.}

\mac{The numerical exploration of plate tectonics has developed rapidly in the past several decades. \cite{gurnis1988large} provided the first time-dependent numerical simulations of continental drift, where multiple continents were allowed to merge and diverge. As in this work, many other numerical and theoretical endeavors \citep{zhong1993dynamic,lowman1993mantle,lowman1995mantle,lowman1999thermal,lowman1999effects,lowman2001influence,zhong2000role} employ geophysical parameters of the mantle and enable rapid advancement of our understanding of the interior of Earth. It has also become clear that the two-way coupling between the continental plates and the mantle convection results in diverse dynamics \citep{gurnis1988large,zhong1993dynamic,phillips2005heterogeneity,Whitehead2015}. Most notably, large plates are observed to have more consistent motions, while small plates tend to move sporadically \citep{gurnis1988large,Whitehead2015}. These observations were recently examined by \cite{mao2019dynamics,mao2021insulating} through resolved numerical simulations. } 

\mac{The aforementioned works confirm that continental plates are not only passive to the mantle flow advection beneath, but are also affecting the flow structure through the \textit{thermal blanket effect}: The continental crust is known to have a much lower heat flux compared to the oceanic crust due to its large crust depth \citep{mao2021insulating}, so the continental plates essentially serve as a blanket that prevents heat from escaping and warms up the mantle beneath. }\nick{The warm and light mantle tends to rise, forming an upward convective flow. As this flow moves towards the surface of Earth it diverges and creates a fluid forcing beneath the continental plate, transporting the plate away. This is the current understanding of continental drift, and the thermal blanket effect has been verified both numerically and experimentally \citep{gurnis1988large,Zhong2005,zhong2007a,zhong2007b}. }

\mac{This manuscript aims to provide a new angle for modeling plate tectonics through a low-dimensional model. After conducting direct numerical simulations (DNS) of the plate-flow interaction in a 2-dimensional periodic domain, we propose a stochastic model of the plate motion and show how the moving plate mechanically and thermally couples to a convecting fluid flow beneath it. Only retaining the most basic physics of thermal convection, this model recovers the dynamics observed in the full DNS and captures the transition of plate dynamics seen in \cite{gurnis1988large,Whitehead2015,mao2019dynamics,mao2021insulating}. }

\mac{In what follows, we summarize the equations and numerical methods in \cref{sec-num}, and present the numerical results in \cref{sec-num-results}. The stochastic model will be systematically derived in \cref{sec-stochastic}, and its application to the convection domain with various aspect ratios will be discussed in \cref{sec-aspect}. Finally, we summarize and discuss our results in \cref{discussion}.}

\section{Numerical model}
\label{sec-num}

\begin{figure*}
 \includegraphics[width=0.55\textwidth]{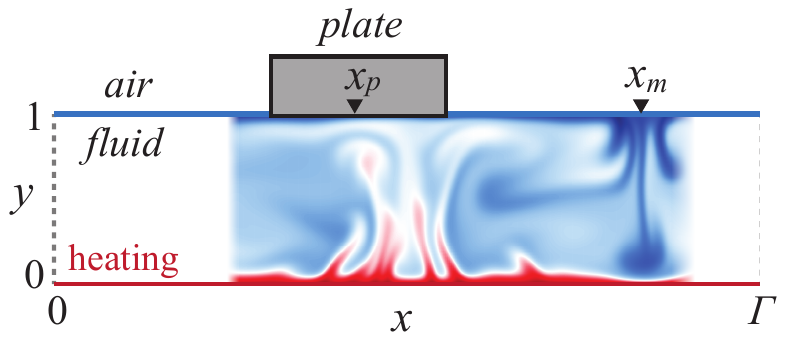}
 \centering
 \caption{Schematics of the moving plate and convecting fluid. The fluid domain is bounded between $y\in(0,1)$ and periodic in $x$, and the floating plate of width $d$ has center location $x_p$.}
\label{fig1}
\end{figure*}
\subsection{Flow equations}
The configuration of our numerical simulation is shown in \cref{fig1}, where a solid plate centered at location $x=x_p$ floats on top of a convecting fluid that is bounded in the $y$ direction and periodic in the $x$ direction. Throughout this study, all lengths are rescaled by the fluid depth $H$, time is rescaled by the diffusion time $H^2/\kappa$ ($\kappa$ is thermal diffusivity), and temperature is rescaled by the temperature difference $\updel T$ between the bottom and top free surface. The $x$ direction of the fluid domain is periodic with period $\Gamma = D/H$ ($D$ is the domain width), so the overall computational domain is $x\in(0,\Gamma)$ and $y\in(0,1)$ as shown in \cref{fig1}. With the Boussinesq approximation, the resulting PDEs for flow speed $\mathbf{u} = (u,v)$, pressure $p$, and temperature $\T \in [0,1]$ are
\begin{align}
    &\frac{D\mathbf{u}}{Dt} = -\nabla p + \Pra{}\,\Lap\mathbf{u} + \Ra{}\,\Pra{}\, \T\, \mathbf{e}_y,\label{ns-eqn}\\
    &\nabla\cdot \mathbf{u} = 0,\label{incompress}\\
    &\frac{D \T}{Dt} = \Lap \T.\label{heat-eqn}
\end{align}
Here, the Rayleigh number is $\Ra{} = \alpha g \updel T H^3 /\nu\kappa$ and the Prandtl number is $\Pra{} = \nu/\kappa$, where $\nu, \alpha$ and $g$ are the kinematic viscosity, the thermal expansion coefficient of the fluid, and the acceleration due to gravity. Simple modifications to the flow solver can be adapted for the geophysical mantle convection, but as we wish to consider a more general case of fluid-structure interactions and to apply our theory to future laboratory experiments, we preserve the inertia of both the fluid and the solid plate in this study. 

\subsection{Boundary conditions}
Without the presence of a plate, the boundary conditions are straightforward: the flow velocity $\uu = (u,v)$ is no-slip at the fluid/solid boundary and shear-free at the air/fluid interface; The temperature $\theta$ is 1 at the bottom and 0 at the air/fluid boundary.

This yields the boundary conditions for the bottom surface $y=0$, 
\begin{equation}
    \T = 1,\,  u=v=0 \quad \mbox{at  } y=0. \label{bottomBC-uv}
\end{equation}

At the top surface, the plate is effectively shielding the heat from escaping, we thus take $\theta_n =0$ there while set the flow to be no-slip with respect to the moving plate. The resulting boundary conditions are
\begin{align}
    \T = 0,\,  u_y=v=0 \quad &\mbox{for } y=1 \mbox{ and } x\notin P,\\
    \T_y = 0,\,  u = u_p, v=0 \quad &\mbox{for } y=1 \mbox{ and } x\in P. \label{topBC}
\end{align}

Alternatively, these conditions can be enforced as
\begin{equation}
\label{movingBC-uv}
  \begin{cases}
    (1-\mathbb{1}_{P})\,\T + \mathbb{1}_{P}\,\T_y = 0\\
    \mathbb{1}_{P}\,u+(1-\mathbb{1}_{P})\,u_{y} = u_p\quad\quad\mbox{at } y=1, \\
    v = 0
\end{cases}  
\end{equation}
where $\mathbb{1}_{P}$ is an indicator function that takes the value of 1 under the plate and 0 otherwise.

\subsection{Plate dynamics}
The fluid shear force directly drives the plate motion, so 
\begin{equation}
    m \dot{u}_p = -\Pra{} \int_P \pd{u}{y}(x,1,t) dx. \label{plate-dynamics}
\end{equation}
Here $u_p = \dot{x}_p$ is the plate velocity, $m = \rho d$ is the dimensionless mass of the plate with linear density $\rho$ and width $d$, and the integration area $P = \{x\, |\, x\in(x_p-d/2, x_p+d/2)\}$ is the region under the plate. 

\subsection{Parameters and numerical method}
The numerical method solving \cref{ns-eqn,incompress,heat-eqn} with \cref{bottomBC-uv,movingBC-uv,plate-dynamics} is detailed in Appendix A, where we use a Fourier-Chebyshev spectral method to obtain resolved and accurate numerical solutions. In all simulations, we choose $\Ra{} = 10^6$, $\Pra = 7.9$, $\Gamma = 1-16$, and $m = 4d$ ($d$ is the plate width), matching the parameters of water convection in experiments \citep{zhang2000periodic,Zhong2005}.

\section{Numerical results}
\label{sec-num-results}
\begin{figure*}
 \includegraphics[width=0.8\textwidth]{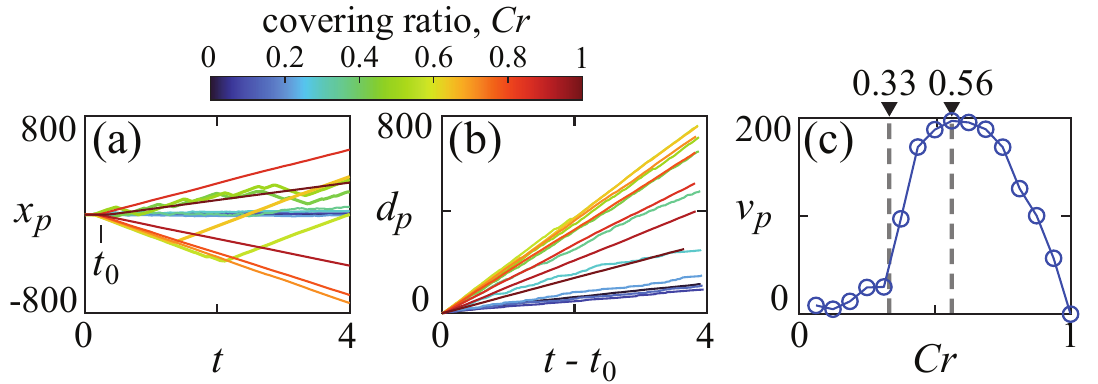}
 \centering
 \caption{Motion of a plate floating on top of a convective fluid. (a) Trajectories of plates of various sizes. The covering ratio is $\Cr{} = d/\Gamma$ and the time $t_0$ marks the beginning of plate motion. (b) The total displacement of the plates, where $d_p$ is defined through $\dot{d}_p = |\dot{x}_p|$. (c) Average plate speed $v_p = \langle|\dot{x}_p|\rangle$ becomes high when $\Cr>0.33$ and reaches a maximum at $\Cr{} = 0.56$. In these simulations, $\Gamma = 4$, $\rho=4$, $\Ra{} = 10^6$, and $\Pra{} = 7.9$.}
\label{fig2}
\end{figure*}

Several trajectories of plates with various sizes are shown in \cref{fig2}(a), and we can immediately see how the plate's size affects its motion. \mac{We define the \textit{covering ratio} $\Cr{} = d/\Gamma$ to measure how much of the free surface is covered by the plate of width $d$, where the fluid aspect ratio is $\Gamma = 4$ for all the results in this section.} For small plates, their net displacement is small, which can be better seen from the total displacement $d_p(t) = \int_0^t |u_p(t')|\,dt'$ shown in \cref{fig2}(b).  Increasing the plate size, linear motion appears as $\Cr{}$ becomes greater than 0.33, as seen in the green trajectories in \cref{fig2}(a). These trajectories are subject to reversals, as there is an effective noise from the turbulent fluid forcing. As $\Cr{}$ further increases, the linear motion becomes more persistent, as the reversals of plate motion become rare when $\Cr{}\to1$ in \cref{fig2}(a). We note that similar dynamical behaviors have been seen in geophysical Stokes flow simulations \citep{mao2021insulating}, therefore the coupling mechanism between the moving plate and flow beneath must be similar for different flow regimes.

From the total displacement $d_p$, one can see that a maximum plate speed is achieved at around $\Cr{}\approx 0.5$, and this can be confirmed by plotting the time-averaged plate speed $v_p = \langle|\dot{x}_p|\rangle$ in \cref{fig2}(c). The average velocity $v_p$ remains low for small plates, but increases significantly for $\Cr{}>0.33$ and reaches a maximum around $\Cr = 0.5$.

\begin{figure}
 \includegraphics[width=1\textwidth]{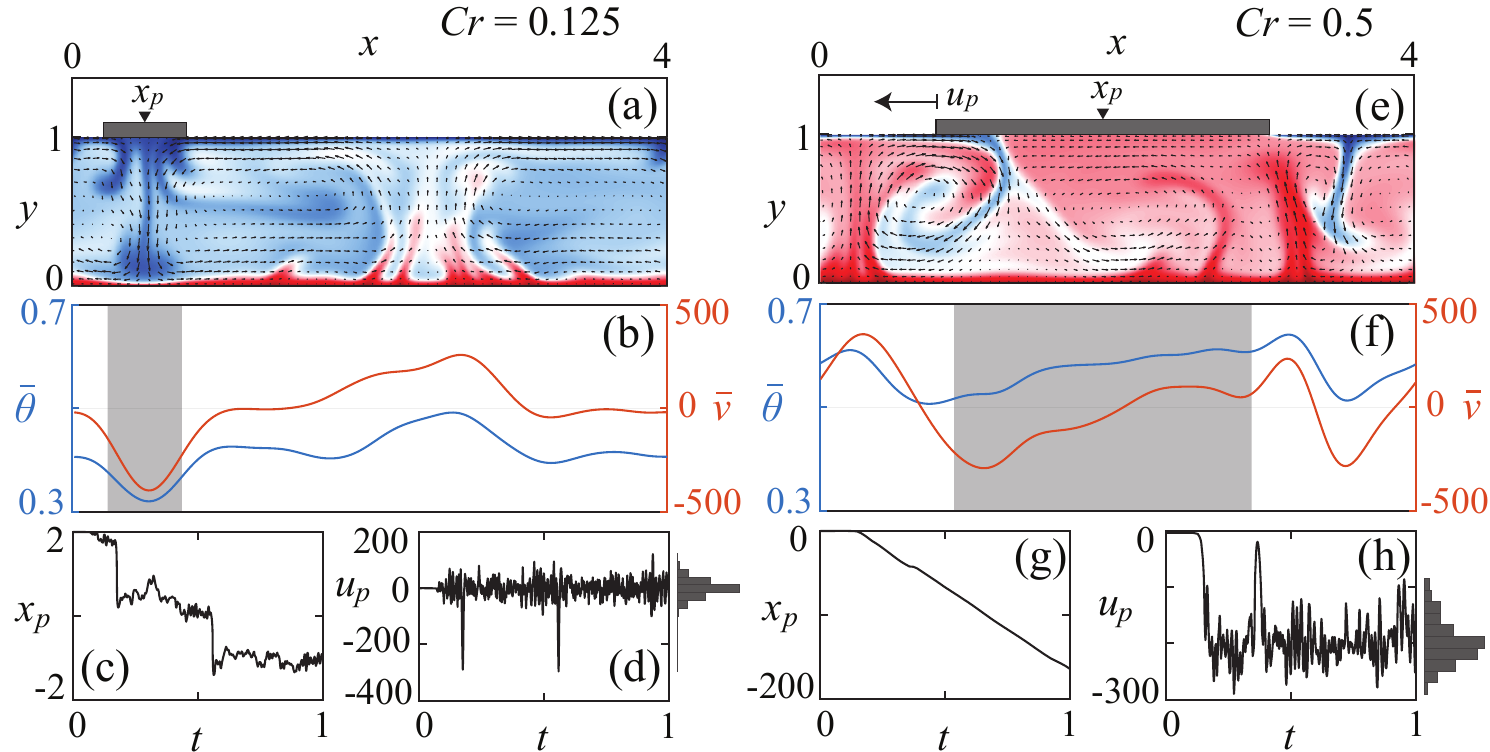}
 \centering
 \caption{Dynamics of floating plate with $\Cr{} = 0.125$ [(a)-(d)] and $\Cr{} = 0.5$ [(e)-(h)]. (a) A typical snapshot of the flow and temperature distributions under the plate with $\Cr{} = 0.125$. (b) The $y$-averaged temperature and vertical flow velocity corresponding to (a), the shaded area indicates the position of the plate. (c) The plate displacement is a random process. (d) Plate velocity is a random variable with mean 0, whose PDF is a Gaussian distribution as indicated by the histogram.  (e) At $\Cr{} = 0.5$, typical flow and temperature distributions showing the plate is transported by the surface flow. (f) The $y$-averaged temperature and vertical flow velocity corresponding to (e). (g) The plate displacement is linear in time, indicating a unidirectional translation. (h) Plate velocity has nonzero mean. Supplementary movies of these simulations are included.}
\label{fig3}
\end{figure}

To investigate the transition between dynamical states, the typical flow and temperature distributions in the fluid are shown in \cref{fig3}. In \cref{fig3}(a), a small plate with $\Cr{} = 0.125$ is placed on the convecting fluid and it is attracted by the center of downwelling fluid at $x_m$ [\cref{fig1}], where the surface flow forms a sink. This sink is a stable equilibrium for the plate, as any deviations from this sink will result in a restoring fluid force acting on the plate. The structure of this flow sink can be further seen in \cref{fig3}(b), where both the $y$-averaged temperature $\bar\theta = \int_0^1\theta\, dy$ and the $y$-averaged vertical flow velocity $\bar v = \int_0^1 v\, dy$ reach their minima.

Following this surface flow pattern, the plate displacement $x_p$ is stochastic as shown in \cref{fig3}(c). Due to the random forcing from turbulent flows, the plate location is subject to noise that can be seen affecting the plate velocity $u_p$ in \cref{fig3}(d), whose histogram shows a Gaussian distribution. It is rare but not impossible for the plate to experience a strong ``wind" from the flow, which can push the plate away from the flow sink, across the flow source, and back to the sink again, resulting in the jumps in $x_p$ seen in \cref{fig3}(c).

\Cref{fig3}(e) shows the dynamics of a plate with $\Cr{} = 0.5$. In this case, the plate motion is unidirectional as shown in \cref{fig3}(g)-(h), with velocity $u_p$ that has a nonzero mean. Shown in \cref{fig3}(e)-(f), the moving plate tends to situate between the flow sink and source. As the surface flow pushes the plate towards its sink, the distribution of flow temperature also shifts, leading to a moving plate chasing a moving surface flow sink.

This is a direct consequence of the thermal blanket effect: When the plate is large enough, the temperature increases beneath it as heat cannot escape there. This local warming modifies the flow temperature and effectively pushes the cold, downwelling fluid away, resulting in a shift of the flow sink location. Overall, the plate moves towards the cold flow sink while simultaneously pushing the sink away. Thus a simple dynamics exists for this seemingly complicated fluid-structure interaction problem, and we will derive a model from these observations.

\section{Stochastic model}
\label{sec-stochastic}

As seen in \cref{fig3}, the variation of the $y$-averaged temperature $\bar{\theta}$ strongly affects the flow pattern. To capture the variational and periodic nature of $\bar\theta$, we approximate it with its lowest nontrivial Fourier mode,
\begin{equation}
\label{temp}
    \bar{\theta}(x,t) = \alpha - \alpha\cos{r[x-x_m(t)]},
\end{equation}
where $x_m$ is the location of surface flow sink in \cref{fig1}, $r = 2\pi\Gamma^{-1}$ is the wavenumber, and the constant $\alpha$ measures the strength of temperature variation. 

Induced by this temperature distribution, the surface flow velocity $U(x,t) = u(x, 1, t)$ can be approximated as
\begin{equation}
\label{surfu}
    \surf(x,t) = -\beta\sin{r[x-x_m(t)]},
\end{equation}
where $\beta>0$ is the surface flow strength. Indeed, this surface flow profile has a sink at $x = x_m$: Small deviations from $x_m$ results in $\surf>0$ for $x<x_m$ and $\surf<0$ for $x>x_m$, so the flow locally points towards $x = x_m$.

We note that this surface flow profile does not match the plate velocity at the solid/fluid boundary, and the mismatch between $\surf$ and $u_p$ allows us to estimate $\partial u/\partial y(x,1,t)$ and the resulting shear stress, leading to the plate acceleration
\begin{equation}
\label{updot1}
    \dot{u}_p(t) = -\frac{\Pra}{m} \int_P \frac{u_p(t)-\surf(x,t)}{\delta}\, dx + \sigma \dot{W}(t).
\end{equation}
Here, $\delta$ is the momentum boundary layer thickness \citep{schlichting2016boundary} that is determined by $\Ra{}$ and $\Pra{}$, so we assume it to be constant in our study. We also include a white noise with standard deviation $\sigma$, representing the turbulent fluid forcing. \mac{Overall, \cref{surfu,updot1} represent a flow profile that has both the large-scale circulations and the small-scale turbulent flows, consistent with observations of high-$\Ra{}$ thermal convection \citep{Ahlers2009,Huang2022a}.}

To model the moving plate as a thermal blanket, we look at the $y$-averaged heat equation, 
\begin{equation}
    \pd{\bar{\theta}}{t} = \ppd{\bar{\theta}}{x} + q(x,t).
\end{equation}
Here we have ignored the flow advection, and $q(x,t) = \pd{\theta}{y}(x,1,t)-\pd{\theta}{y}(x,0,t)$ is the vertical heat flux passing through location $x$. Assuming the heat leaving the fluid-air interface obeys Newton's law of cooling and no heat penetrates the plate, we can rewrite the heat equation as 
\begin{equation}
\pd{\bar{\theta}}{t} = \ppd{\bar{\theta}}{x} - \gamma \bar{\theta}(1-\indp).
\end{equation}
The indicator function $\indp(x)$ is 1 when $x\in P$ and 0 otherwise, and the constant $\gamma$ models the rate of cooling. We now plug in the value of $\bar\theta$ from \cref{temp} and integrate over $x$, which leads to an ODE for $x_m$,
\begin{equation}
\label{xmdot}
    \dot{x}_m(t) = \frac{\gamma}{\pi} \int_P[1-\cos r(x_m-x)]\sin{r(x_m-x)}\, dx.
\end{equation}

The integrals in \cref{updot1,xmdot} can be evaluated exactly. Defining a phase angle $\phi = r(x_p-x_m)$, we arrive at a closed dynamical system for $(u_p, \phi)$,
\begin{align}
    \dot{u}_p &= -\frac{\beta\lambda}{\pi\Cr}\sin(\pi\Cr) \sin\phi - \lambda u_p + \sigma \dot{W},\label{udot}
\\
    \dot{\phi} &= ru_p + \frac{2\gamma}{\pi}\sin(\pi\Cr) \sin\phi-\frac{\gamma}{2\pi}\sin(2\pi\Cr) \sin(2\phi),\label{phidot}
\end{align}
where $\lambda = \Pra/(\rho\delta)$. Once the dynamics of $(u_p, \phi)$ is known, the dynamics of $(x_p, x_m)$ can be calculated through $\dot{x}_p = u_p$ and $x_m = x_p-r^{-1}\phi$.

\begin{figure}
 \includegraphics[width=\textwidth]{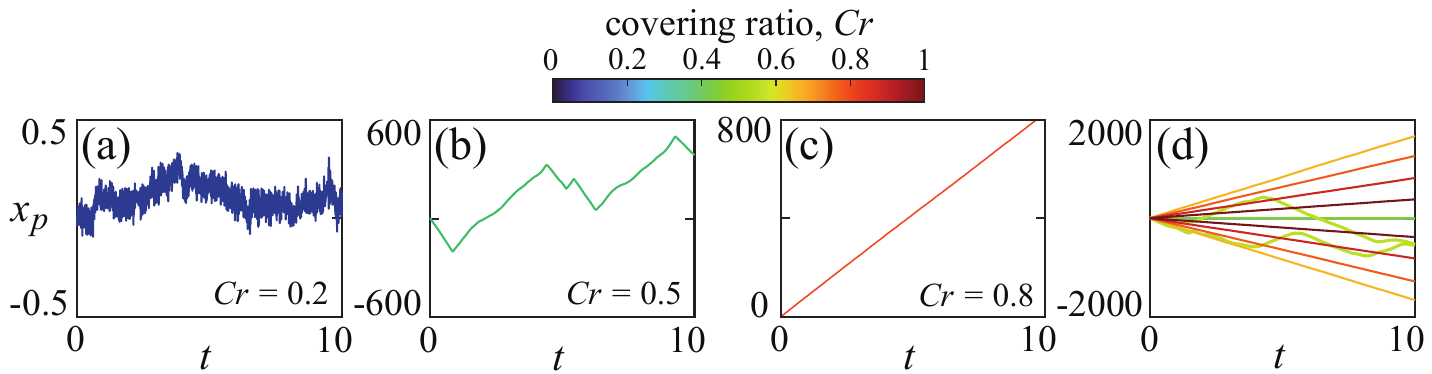}
 \centering
 \caption{Simulated trajectories from the stochastic model. (a) The dynamics of the small plate is a random walk. (b) The medium-sized plate has a nonzero transnational velocity whose direction is subject to reversals. (c) The translational motion of the large plate is more persistent. (d) Trajectories of all simulated paths, whose dynamics recover \cref{fig2}(a).}
\label{fig4}
\end{figure}

There are four parameters in this model: $\beta$ as the strength of surface flow, $\lambda=\Pra/(\rho\delta)$ as a damping coefficient, $\gamma$ as the rate of surface cooling, and $\sigma$ as the random fluid forcing. Physically, the surface cooling rate $\gamma$ is affected by the surface flow strength $\beta$, so we take $\gamma = r \beta$ in this model which results in the correct dynamics. The remaining parameters can be estimated from the numerical simulations, and their values and estimation procedures are \mac{included in Appendix B}.

Starting with $\phi(0) = 0$ and a random value of $u_p(0)$, \cref{fig4}(a)-(c) show some typical trajectories of $x_p(t)$ at different $\Cr{}$. In \cref{fig4}(a), the trajectory of the small plate ($\Cr{} = 0.2$) is noise-driven. For the medium plate of $\Cr{} = 0.5$, \cref{fig4}(b) shows its trajectory is composed of linear translations with reversals. For the large plate of $\Cr{} = 0.8$, \cref{fig4}(c) indicates that the translation is unidirectional. \mac{The plate speed in \cref{fig4}(c) is comparable to the full DNS result in \cref{fig2}, and this speed decreases as $\Cr$ further increases. The typical displacement $x_p$ for plates with various sizes is shown in \cref{fig4}(d)}, which resembles \cref{fig2}(a) and has a transition between the noise-driven and linear motions at $\Cr{}\approx0.3$. Thus, this simple model captures all the key features of the full numerical simulation.

Without noise, the critical behavior of the dynamical system \cref{udot,phidot} can be further analyzed. For small $\Cr{}$, it is easy to see that \cref{udot,phidot} have $u_p =0$, $\phi = 2\pi N$ (where $N$ is an integer) as equilibria, which are stable and reflect the \textit{passive} state of plate motion. Increasing $\Cr{}$, new equilibria appear at $\Cr{}^* = 1/3$. For $\Cr{}>\Cr^*$, it becomes possible to have a nonzero plate velocity $u_p^* = (\beta/\pi)\hat{u}_p^*$, where
\begin{equation}
\label{upstar}
    \hat{u}^*_p = -\frac{\sin\pi\Cr}{\Cr}\sin\phi^*
\end{equation}
and the equilibrium $\phi^*$ can be determined from
\begin{equation}
\label{phistar}
    \cos{\phi^*} = \frac{1-(2\Cr)^{-1}}{\cos\pi\Cr}.
\end{equation}
These states represent the \textit{translation} of the plate. We note that \cref{upstar,phistar} are only functions of $\Cr{}$, and thereby independent of all other parameters assumed in this model. To recover the dimensional plate velocity $u_p^*$, one only needs to know the flow speed factor $\beta/\pi$. 

\begin{figure}
 \includegraphics[width=0.6\textwidth]{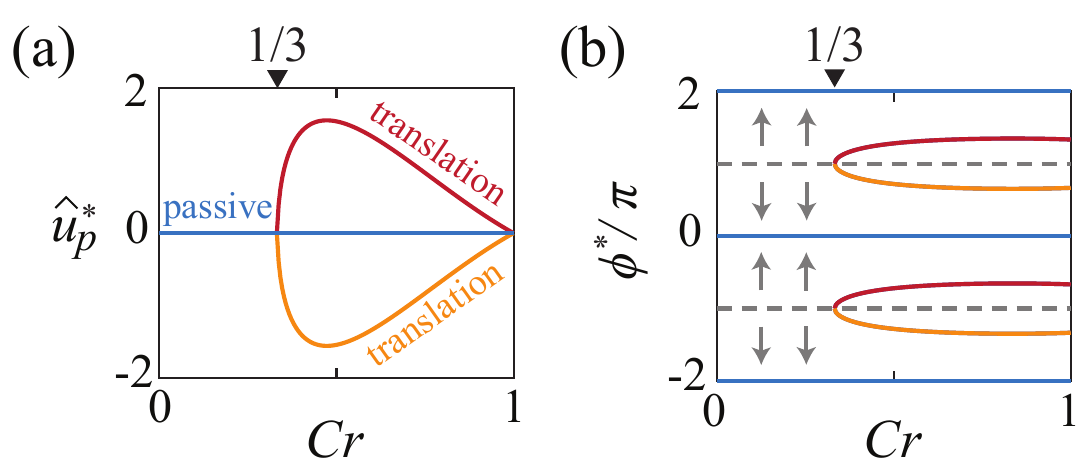}
 \centering
 \caption{Equilibrium values of the plate velocity and phase angle. (a) Dimensionless plate velocity $\hat{u}_p$ is 0 for small $\Cr$, but becomes translational as $\Cr$ increases. (b) The phase angle $\phi$. Blue lines represent the passive state where the plate is attracted by the flow sink, and red/orange curves show the equilibrium phase of the translational state. The surface flow direction is labelled with arrows. }
\label{fig5}
\end{figure}

The possible values of $\hat{u}_p^*$ and $\phi^*$ are shown in \cref{fig5}. For $\Cr{}<\Cr^* = 1/3$, $u^*_p = 0$ and $\phi^* = 2\pi N$ are the only possible equilibria which reflect the passive nature of the small plate that is always attracted by the surface flow sink. For $\Cr{}>\Cr^*$, new phases appear as $\phi^* = (2N+1)\pi\pm \arccos{\left([(2\Cr)^{-1}-1](\cos\pi\Cr)^{-1}\right)}$, which are solutions of \cref{phistar} and become stable for large $\Cr{}$. They indicate that the larger plate tends to sit between the surface flow sink ($\phi=2N\pi$) and source [$\phi=(2N+1)\pi$], confirming our observations in \cref{fig3}. As the surface flow points from its source to its sink [see arrows in \cref{fig5}(b)], these new phases indicate two possible plate velocities that are given by \cref{upstar} and shown in \cref{fig5}(a). Furthermore, \cref{fig5}(a) resembles \cref{fig2}(c) as the plate velocity vanishes for $\Cr\to\Cr^*$ and $\Cr\to 1$, and obtains its maximum around $\Cr = 0.5$.

\mac{Through this simple model, we see clear physics of how the solid plate interacts with the fluid beneath, and the covering ratio $\Cr{}$ serves as a measure of the strength of the thermal blanket effect. For small $\Cr{}$, the thermal blanket effect is weak thus the plate motion is passive to the fluid. Increasing $\Cr{}$ beyond $\Cr^*$, the thermal blanket effect is strong enough to alter the flow and temperature distributions, generating an upwelling that can lead to plate motion. Both \cref{fig2}(c) and \cref{fig5}(a) suggest that the plate speed peaks at $\Cr{}\approx 0.5$ and vanishes as $\Cr{}\to1$, which also reflects the competition between the thermal blanket effect and the flow convection. As $\Cr{}$ increases, more area of the free surface is covered by the plate and the fluid force beneath is averaged in a larger domain. As this domain may cover both the upwelling and downwelling flows, the total fluid force is affected by $\Cr$. This can be seen in \cref{updot1}, where the surface flow velocity $U$ provides plate acceleration. In \cref{surfu}, $U$ is modelled as sinusoidal and this profile is integrated in \cref{updot1}, therefore increasing the integration area in \cref{updot1} to half of the open surface ($\Cr = 1/2$) will cover the highest contribution of $U$. Further increasing the covering area will thereby decrease the contribution of $U$, as the integration domain is more than half a period of the sine function. In an extreme case, $\Cr = 1$ indicates an integration of $U$ for a full period, leading to zero fluid force as seen in \cref{fig5}(a).}

\section{Aspect ratio effect}
\label{sec-aspect}

All the results discussed so far focus on the convective fluid domain of aspect ratio $\Gamma = 4$, which matches the geometry of many experimental investigations. Varying the domain aspect ratio will certainly affect the dynamics of the convecting fluid and moving plate, as a more complicated, multi-roll flow structure emerges \citep{Ahlers2009}. In this section, we investigate the plate dynamics at $\Gamma =2$, 4, 8, and 16 while keeping other dynamical parameters the same as described in \cref{sec-num,sec-num-results,sec-stochastic}.

\begin{figure}
 \includegraphics[width=\textwidth]{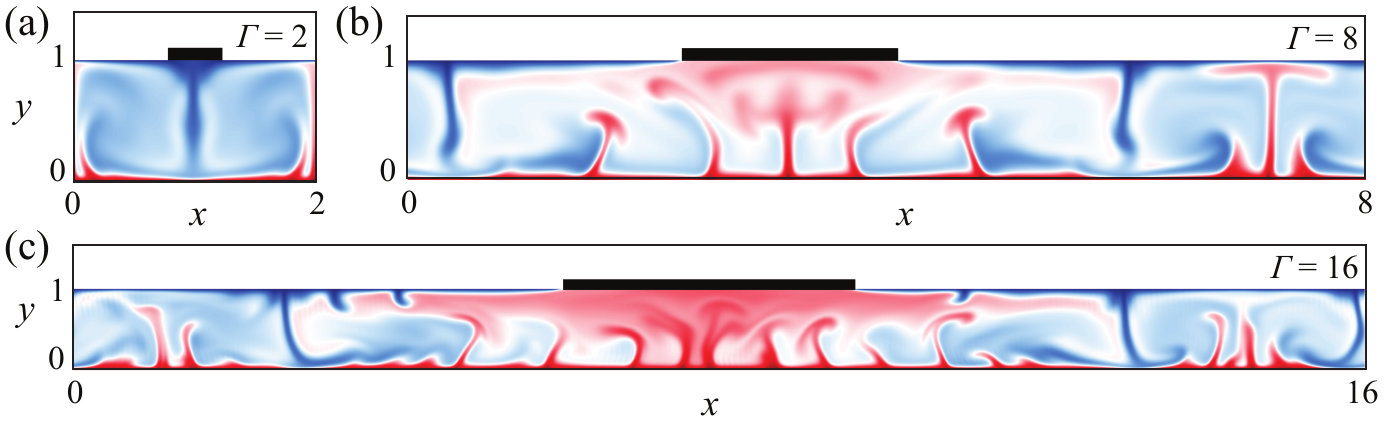}
 \centering
 \caption{ \mac{Snapshots of the convective fluid at various aspect ratio $\Gamma$. (a) $\Gamma = 2$. (b) $\Gamma = 8$. (c) $\Gamma = 16$. In these simulations, all  other parameters are fixed at $\Cr{} = 0.2$, $\rho=4$, $\Ra{} = 10^6$, and $\Pra{} = 7.9$.}}
\label{fig6}
\end{figure}

\begin{figure}
 \includegraphics[width=\textwidth]{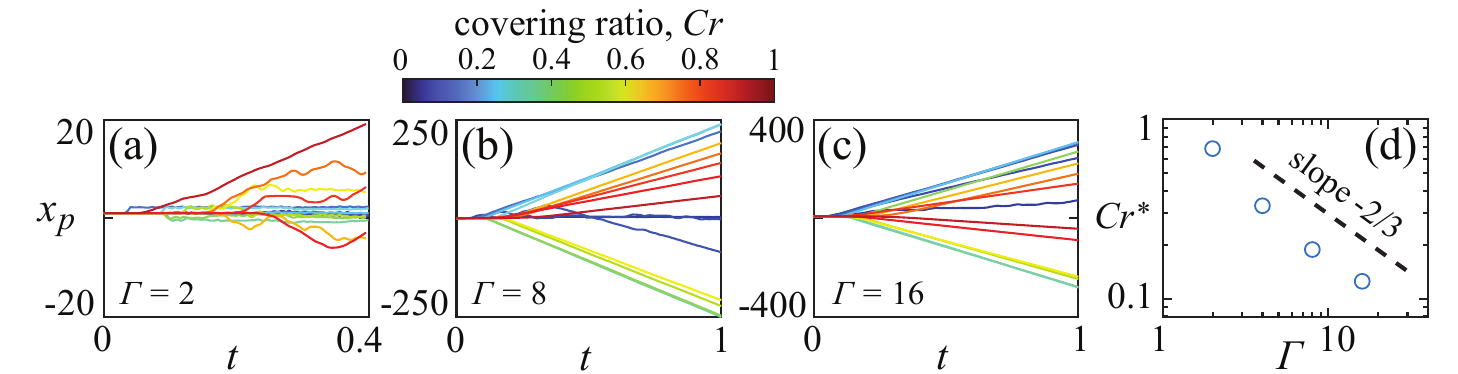}
 \centering
 \caption{\mac{Simulated trajectories from the DNS at the aspect ratio of (a) $\Gamma = 2$, (b) $\Gamma = 8$, and (c) $\Gamma = 16$. In these simulations, $\rho=4$, $\Ra{} = 10^6$, and $\Pra{} = 7.9$, as in \cref{fig2,fig3}. (d) The critical $\Cr^*$ separating the passive and translating states has a -2/3 power-law scaling with $\Gamma$.}}
\label{fig7}
\end{figure}

\Cref{fig6} shows some typical temperature distributions of DNS results at $\Gamma = 2-16$. Clearly, a more complicated flow structure emerges as more convection cells appear with increasing $\Gamma$. \Cref{fig7} shows trajectories of the plate at different $\Gamma$; a common feature reemerges as the small plate moves little while the large plate translates.

To extend our simple model to cases of high aspect ratio, we consider a bulk temperature and surface flow profile with a more complicated spatial dependence, 
\begin{align}
    \bar{\theta}(x,t) &= \alpha - \alpha\cos{kr[x-x_m(t)]},\\
    \surf(x,t) &= -\beta\sin{kr[x-x_m(t)]},
\end{align}
where the integer $k$ is the most dominant wave number in the Fourier spectrum. The inclusion of wave number $k$ is inspired by the fact that the convection might have a multi-roll structure, so the temperature and flow profiles above indicate that there are $2k$ convection rolls in the fluid with $k$ surface sinks and $k$ surface sources. We track one of the surface sinks $x_m(t)$, which is also the location of lowest fluid temperature. 

Through similar derivations as in \cref{sec-stochastic}, we can again obtain a stochastic dynamical system for the phase $\phi = r(x_p - x_m)$ and the plate velocity $u_p$,
\begin{align}
    \dot{u}_p &= -\frac{\beta\lambda}{k\pi\Cr}\sin(k\pi\Cr) \sin(k\phi) - \lambda u_p + \sigma \dot{W},\label{ukdot}
\\
    \dot{\phi} &= ru_p + \frac{2\gamma}{k^2\pi}\sin(k\pi\Cr) \sin(k \phi)-\frac{\gamma}{2k^2\pi}\sin(2k\pi\Cr) \sin(2k \phi).\label{phikdot}
\end{align}

Without noise, it is easy to verify that \cref{ukdot,phikdot} have passive equilibria $u_p^* = 0$ and $\phi^* = 2m\pi/k$, where $m$ is any integer. The Jacobian of such passive states is 
\begin{gather}
 J =
  \begin{bmatrix}
  -\lambda & -\beta\lambda(\pi \Cr)^{-1}\sin(k\pi\Cr)
   \\
   r & 2\gamma(k\pi)^{-1} \sin(k\pi\Cr)(1-\cos(k\pi\Cr))
   \end{bmatrix}.
\end{gather}

Evaluating the trace and determinant of $J$ in the limit of $\Cr\to0$ yields
\begin{align}
    \mbox{tr}(J) &\to -\lambda <0,\\
    \mbox{det}(J) &\to kr\beta\lambda > 0.
\end{align}
Therefore both eigenvalues of $J$ are negative, indicating stable passive states for small $\Cr{}$ and confirming our numerical observations.

The stability of passive states is lost when one or both eigenvalues of $J$ become positive, therefore $\mbox{det}(J) = 0$ provides an equation to determine the critical $\Cr^*$. After simplification, we have
\begin{equation}
    \frac{\Cr^*}{k}(1-\cos k\pi \Cr^*) = \frac{\beta r}{2 \gamma}.
    \label{cr-eqn}
\end{equation}

The root of \cref{cr-eqn} can be determined numerically, given that the parameters $\beta$ and $\gamma$ are known. In laboratory experiments and geophysical plate tectonics, the surface flow speed scale $\beta$ and ventilation coefficient $\gamma$ can be properly estimated and used to determine $\Cr^*$, therefore \cref{cr-eqn} offers the possibility of determining plate mobility through physical parameters. 

We note that $\Cr^*$ is usually small for large $\Gamma$ as shown in \cref{fig7}, and the root of \cref{cr-eqn} in this limit can be shown as 
\begin{equation}
    \Cr^*\approx\left(\frac{\beta r}{\pi^2 \gamma k}\right)^{1/3} \sim \Gamma^{-2/3}.
    \label{cr-approx}
\end{equation}
Here we have used the relations $r \sim \Gamma^{-1}$ and $k\sim \Gamma$, with the latter indicating the number of convection rolls is proportional to the aspect ratio $\Gamma$. \Cref{cr-approx} can indeed be verified-- as shown in \cref{fig7}(d), the critical $\Cr^*$ measured from DNS data does follow the -2/3 power law with $\Gamma$. 

Much more can be investigated through the dynamics of \cref{ukdot,phikdot}, and future experimental studies can certainly be used to further address the interaction between the plate and convective fluid below.

\section{Discussion}
\label{discussion}

In this work, we have numerically and theoretically explored the mechanical and thermal coupling between a moving plate and a convecting fluid beneath it. \mac{Inspired by present and past works \citep{gurnis1988large,Whitehead2015,mao2019dynamics,mao2021insulating}, we propose a stochastic model showing that the plate size (covering ratio) is a deciding factor for the strength of the thermal blanket effect. For small plates, the shielding effect is weak and the plate motion is passive to the flow structure; For large plates, the flow beneath becomes warm enough and an upwelling center is formed, pushing the plate away and resulting in its translation. The proposed simple model consists of minimal assumptions about the flow and its mechanical and thermal coupling to the plate, however it is capable of predicting the dynamical transition of the plate motion. Although the DNS is conducted at a parameter regime similar to laboratory experiments \citep{Zhong2005,zhong2007a,zhong2007b}, this stochastic model is Reynolds number independent, therefore allowing it to be applied to both laboratory and geophysical scenarios.}

\mac{Laboratory-scale experiments are usually conducted in a bounded convection system, therefore the plate is limited to move between two walls. The flow structure and its coupling to the plate is very different in bounded convection, as previous works show the flow has two counter-rotating large-scale circulations, whose strengths are modulated by the location of the plate \citep{Zhong2005,zhong2007a,zhong2007b}. The plate is also bounded by the solid boundaries, so it stops moving once touching the wall. Modeling the plate dynamics in this case is not trivial, and advanced tools such as stochastic variational inequalities \citep{mac2018stochastic} can serve as a mathematical means of analyzing such plate-boundary interactions. A periodic domain of convecting fluid is necessary to experimentally verify our stochastic model, and such an experiment would also resemble the mantle convection more closely. Future experiments consisting of an annular convection domain could provide more details to validate the stochastic model.}

Although here we only investigate the dynamics of a single plate, our numerical method is capable of handling multiple plates providing that their interactions can be properly modeled. We are currently investigating such interactions, which has led to even more diverse and unpredictable dynamics. For example, if two small plates each have $\Cr{}<1/3$ but their combined size reaches $\Cr{}>1/3$, we have seen that each individual plate moves randomly but their combined ``super continent" can translate. In the geophysical case of plate tectonics, plate interactions are the reason for many volcanic activities and mountain formations, therefore understanding the converging and diverging motion of nearby plates might offer new insights into the fluid mechanics behind these geophysical events. 

For simplicity, the simulation and model in this work are both two dimensional, and extending our results to three dimensions is a current priority. Using the Chebyshev-Fourier-Fourier method, we have implemented the numerical solver for evolving a plate sitting on top of a three-dimensional domain that is periodic in two horizontal directions. Moreover, the simulation of plate tectonics on a spherical shell is also possible through the Chebyshev-Chebyshev-Fourier method, which is a configuration closer to the geophysical plate tectonics. Through analyzing the direct numerical simulation results there, we wish to further develop our stochastic model and use it to address the fluid-structure interactions happening inside Earth.

Finally, we note that the geophysical plate tectonics is much more complicated than any experiments or numerical simulations conducted so far, as the interior of Earth is such a complex environment and is still being explored by modern science. But although current simple models cannot fully capture the dynamics of continental drifts, they can hopefully still offer some fluid mechanical insights into the geophysics of Earth. 


\section*{Appendix A: Numerical method}
The Navier-Stokes Boussinesq \cref{ns-eqn,incompress,heat-eqn} in 2D can be written in the vorticity \& stream function format, 
\begin{align}
\frac{D \omega}{Dt}  &= \Pra{}\, \Lap \omega   + \Pra{}\, \Ra{}\, \frac{\partial \T}{\partial x},\label{omegaeq}\\
 -\Lap \psi &= \omega,\ \  \mathbf{u} = \nabla_\perp \psi,\label{psieq}\\
 \frac{D \T}{Dt}  &=  \Lap \T, \label{Teq}
\end{align}
where the $z$-component of vorticity $\omega = \hat{\mathbf{z}}\cdot\nabla\times \mathbf{u}$ and the stream function defined by $\mathbf{u} = \nabla_\perp \psi = (\psi_y, -\psi_x)$ are solved for, alleviating the difficulty of solving for pressure $p$.

In the vorticity \& stream function format, the boundary conditions \cref{bottomBC-uv,movingBC-uv} can be enforced as 
\begin{equation}
    \T = 1,\, \psi=\psi_y=0 \quad \mbox{at  } y=0. \label{bottomBC}
\end{equation}
and
\begin{equation}
\label{movingBC-psi}
  \begin{cases}
    (1-\mathbb{1}_{P})\,\T + \mathbb{1}_{P}\,\T_y = 0\\
    \mathbb{1}_{P}\,\psi_y+(1-\mathbb{1}_{P})\,\psi_{yy} = u_p\quad\quad\mbox{at } y=1.\\
    \psi = 0
\end{cases}  
\end{equation}
In the numerical simulations we soften the edge of indicator function $\mathbb{1}_{P}$, making it smoother in order to reduce numerical error.

The time derivatives in \cref{omegaeq,Teq} are approximated with the \second{} order Adam-Bashforth Backward Differentiation method (ABBD2). At time step $t_n = n\updel T$, we denote $\omega_n(x,y) = \omega(x,y,n\updel T)$, $\psi_n(x,y) = \psi(x,y,n\updel T)$, and $\T_n(x,y) = \T(x,y,n\updel T)$, and \cref{omegaeq,Teq,psieq} become
\begin{align}
\label{omegadisc}
\Lap \omega_n -\sigma_1\omega_n  &= f_n,\\
\label{Tdisc}
\Lap \T_n -\sigma_2 \T_n  &= h_n,\\
\label{psidisc}
 -\Lap \psi_n &= \omega_n,
\end{align}
where 
\begin{align}
\sigma_1 &= \frac{3}{2\,\Pra\,\updel t},\quad \sigma_2 = \frac{3}{2\updel t}, \\[8pt]
f_n &= \Pra^{-1}\left[ 2  (\uu \cdot \grad \omega)_{n-1} - (\uu \cdot \grad \omega)_{n-2}\right]  \label{fn}\\
&\hspace{40pt}- (2\, \Pra\, \updel t)^{-1} \left(4\omega_{n-1}-\omega_{n-2}\right) -\Ra\, \left(\pd{\T}{x}\right)_{n},\notag \\[8pt]
h_n &= \left[ 2  (\uu \cdot \grad \T)_{n-1} - (\uu \cdot \grad \T)_{n-2}\right] - (2\updel t)^{-1} \left(4\T_{n-1}-\T_{n-2}\right). \label{hn}
\end{align}

\Cref{omegadisc,Tdisc,psidisc}, together with the inhomogeneous Robin boundary conditions \cref{bottomBC,movingBC-psi}, are Helmholtz equations that can be solved by standard spectral methods \citep{peyret2002spectral}. More details of this numerical solver will be included in future publications.

Nonlinear terms like $\uu \cdot \grad \T$ and $\uu \cdot \grad \omega$ in \cref{fn,hn} are computed pseudo-spectrally with a simple and efficient anti-aliasing filter \citep{Hou2007}. With given initial and boundary data, \eqref{Tdisc} can be solved first to obtain $\T_{n}$, which is inserted in $f_{n}$ so \eqref{omegadisc} can be solved next. Finally, \eqref{psidisc} is solved with the known $\omega_{n}$.

After solving for the flow and temperature fields, the plate acceleration can be determined as
\begin{equation}
  a_{p,n} = -\frac{\Pra{}}{m} \int_P \ppd{\psi_{n}}{y}(x,1) dx.
\end{equation}

The plate velocity $u_{p,n}$ and plate location $x_{p,n}$ can then be computed through a \second{} order Adam-Bashforth method, 
\begin{align}
x_{p,\,n} &=  x_{p,\,n-1}+\frac{\updel t}{2}( 3u_{p,\,n-1} - u_{p,\,n-2}),\label{loc}\\
u_{p,\,n} &=  u_{p,\,n-1}+\frac{\updel t}{2}( 3a_{p,\,n-1} - a_{p,\,n-2}).\label{spd}
\end{align}

At $\Gamma = 4$, there are typically $256$ Fourier modes in the $x$ direction and $64$ Chebyshev nodes in the $y$ direction, with a time step size of $\updel t = 10^{-6}$. These parameters are tested to yield resolved and accurate numerical solutions.

\section*{Appendix B: Model parameters}

The four parameters involved in the stochastic model can be estimated from the DNS data and auxiliary numerical tests. The detailed procedures are listed below.
\begin{enumerate}
    \item $\beta \approx 400$ is directly measured from the numerical simulation in \cref{fig3}.
    \item $\lambda \approx 200$ is estimated from $\Pra = 7.9$, $\rho = 4$, and $\delta = 0.01$. The boundary layer thickness $\delta$ is estimated from the relation $\delta \sim (2\Nu)^{-1}$, where the Nusselt number $\Nu{}$ is at the order of $10^1$ as measured from the simulation.
    \item $\sigma\approx200$ is estimated from the variance of the plate center $x_p$ for small $\Cr{}$. From \cref{udot}, we have $\sigma^2\approx 2\lambda\mbox{Var}(x_p)$, where $\mbox{Var}(x_p)\approx100$ is measured from the numerical simulation shown in \cref{fig3}.
\end{enumerate}

We note that $\beta$ and $\sigma$ model the surface flow, therefore can be estimated from numerical simulations without floating plate or calculated through scaling relations \citep{Ahlers2009}.

\section*{Supplementary Materials}
Supplementary movies are available at \url{https://math.nyu.edu/~jinzi/research/convectivePlate-1p/Movie}.

\bibliographystyle{tex-library/jfm}
\bibliography{manuscript}

\begin{thebibliography}{32}
\expandafter\ifx\csname natexlab\endcsname\relax\def\natexlab#1{#1}\fi

\bibitem[Ahlers {\em et~al.\/}(2009)Ahlers, Grossmann \& Lohse]{Ahlers2009}
{\sc Ahlers, G., Grossmann, S. \& Lohse, D.} 2009 Heat transfer and large scale
  dynamics in turbulent {R}ayleigh-{B}{\'e}nard convection. {\em Rev. Mod.
  Phys.\/} {\bf 81}~(2), 503.

\bibitem[Elder(1967)]{elder1967convective}
{\sc Elder, J.} 1967 Convective self-propulsion of continents. {\em Nature\/}
  {\bf 214}~(5089), 657--660.

\bibitem[Gurnis(1988)]{gurnis1988large}
{\sc Gurnis, M.} 1988 Large-scale mantle convection and the aggregation and
  dispersal of supercontinents. {\em Nature\/} {\bf 332}~(6166), 695--699.

\bibitem[Hou \& Li(2007)]{Hou2007}
{\sc Hou, T.~Y. \& Li, R.} 2007 Computing nearly singular solutions using
  pseudo-spectral methods. {\em Journal of Computational Physics\/} {\bf
  226}~(1), 379--397.

\bibitem[Howard {\em et~al.\/}(1970)Howard, Malkus \&
  Whitehead]{howard1970self}
{\sc Howard, L., Malkus, W. \& Whitehead, J.} 1970 Self-convection of floating
  heat sources: A model for continental drift. {\em Geophysical and
  Astrophysical Fluid Dynamics\/} {\bf 1}~(1-2), 123--142.

\bibitem[Huang \& Zhang(2022)]{Huang2022a}
{\sc Huang, J.~M. \& Zhang, J.} 2022 Rayleigh{\textendash}{B\'{e}}nard thermal
  convection perturbed by a horizontal heat flux. {\em J. Fluid Mech.\/} {\bf
  954}.

\bibitem[Huang {\em et~al.\/}(2018)Huang, Zhong, Zhang \&
  Mertz]{mac2018stochastic}
{\sc Huang, J.~M., Zhong, J.-Q., Zhang, J. \& Mertz, L.} 2018 Stochastic
  dynamics of fluid--structure interaction in turbulent thermal convection.
  {\em J. Fluid Mech.\/} {\bf 854}.

\bibitem[Kious \& Tilling(1996)]{kious1996dynamic}
{\sc Kious, W.~J. \& Tilling, R.~I.} 1996 {\em This dynamic {E}arth: the story
  of plate tectonics\/}. DIANE Publishing.

\bibitem[Lowman \& Gable(1999)]{lowman1999thermal}
{\sc Lowman, J.~P. \& Gable, C.~W.} 1999 Thermal evolution of the mantle
  following continental aggregation in 3d convection models. {\em Geophysical
  Research Letters\/} {\bf 26}~(17), 2649--2652.

\bibitem[Lowman \& Jarvis(1993)]{lowman1993mantle}
{\sc Lowman, J.~P. \& Jarvis, G.~T.} 1993 Mantle convection flow reversals due
  to continental collisions. {\em Geophysical Research Letters\/} {\bf
  20}~(19), 2087--2090.

\bibitem[Lowman \& Jarvis(1995)]{lowman1995mantle}
{\sc Lowman, J.~P. \& Jarvis, G.~T.} 1995 Mantle convection models of
  continental collision and breakup incorporating finite thickness plates. {\em
  Physics of the Earth and Planetary Interiors\/} {\bf 88}~(1), 53--68.

\bibitem[Lowman \& Jarvis(1999)]{lowman1999effects}
{\sc Lowman, J.~P. \& Jarvis, G.~T.} 1999 Effects of mantle heat source
  distribution on supercontinent stability. {\em Journal of Geophysical
  Research: Solid Earth\/} {\bf 104}~(B6), 12733--12746.

\bibitem[Lowman {\em et~al.\/}(2001)Lowman, King \& Gable]{lowman2001influence}
{\sc Lowman, J.~P., King, S.~D. \& Gable, C.~W.} 2001 The influence of tectonic
  plates on mantle convection patterns, temperature and heat flow. {\em
  Geophysical Journal International\/} {\bf 146}~(3), 619--636.

\bibitem[Mao(2021)]{mao2021insulating}
{\sc Mao, Y.} 2021 An insulating plate drifting over a thermally convecting
  fluid: the effect of plate size on plate motion, coupling modes and flow
  structure. {\em Journal of Fluid Mechanics\/} {\bf 916}, A18.

\bibitem[Mao {\em et~al.\/}(2019)Mao, Zhong \& Zhang]{mao2019dynamics}
{\sc Mao, Y., Zhong, J.-Q. \& Zhang, J.} 2019 The dynamics of an insulating
  plate over a thermally convecting fluid and its implication for continent
  movement over convective mantle. {\em Journal of Fluid Mechanics\/} {\bf
  868}, 286--315.

\bibitem[Meyers {\em et~al.\/}(1987)]{meyers1987encyclopedia}
{\sc Meyers, R.~A. {\em et~al.\/}} 1987 {\em Encyclopedia of physical science
  and technology\/}. Academic Press.

\bibitem[Peyret(2002)]{peyret2002spectral}
{\sc Peyret, R.} 2002 {\em Spectral methods for incompressible viscous flow\/},
  , vol. 148. Springer Science \& Business Media.

\bibitem[Phillips \& Bunge(2005)]{phillips2005heterogeneity}
{\sc Phillips, B.~R. \& Bunge, H.-P.} 2005 Heterogeneity and time dependence in
  3d spherical mantle convection models with continental drift. {\em Earth and
  Planetary Science Letters\/} {\bf 233}~(1-2), 121--135.

\bibitem[Plummer {\em et~al.\/}(2001)Plummer, McGeary, Carlson {\em
  et~al.\/}]{plummer2001physical}
{\sc Plummer, C.~C., McGeary, D., Carlson, D.~H. {\em et~al.\/}} 2001 {\em
  Physical geology\/}. McGraw-Hill Boston.

\bibitem[Schlichting \& Gersten(2016)]{schlichting2016boundary}
{\sc Schlichting, H. \& Gersten, K.} 2016 {\em Boundary-layer theory\/}.
  springer.

\bibitem[Selley {\em et~al.\/}(2005)Selley, Cocks \&
  Plimer]{selley2005encyclopedia}
{\sc Selley, R.~C., Cocks, L. R.~M. \& Plimer, I.~R.} 2005 {\em Encyclopedia of
  geology\/}. Elsevier Academic.

\bibitem[Turcotte \& Schubert(2002)]{turcotte2002geodynamics}
{\sc Turcotte, D.~L. \& Schubert, G.} 2002 {\em Geodynamics\/}. Cambridge
  University Press.

\bibitem[Whitehead {\em et~al.\/}(2011)Whitehead, Shea \&
  Behn]{whitehead2011cellular}
{\sc Whitehead, J., Shea, E. \& Behn, M.~D.} 2011 Cellular convection in a
  chamber with a warm surface raft. {\em Physics of Fluids\/} {\bf 23}~(10),
  104103.

\bibitem[Whitehead(1972)]{Whitehead1972}
{\sc Whitehead, J.~A.} 1972 Moving heaters as a model of continental drift.
  {\em Phys. Earth Planet. In.\/} {\bf 5}, 199--212.

\bibitem[Whitehead(2023)]{Whitehead2022}
{\sc Whitehead, J.~A.} 2023 Convection cells with accumulating crust: Models of
  continent and mantle evolution. {\em Journal of Geophysical Research: Solid
  Earth\/} {\bf 128}~(4), e2022JB025643.

\bibitem[Whitehead \& Behn(2015)]{Whitehead2015}
{\sc Whitehead, J.~A. \& Behn, M.~D.} 2015 The continental drift convection
  cell. {\em Geophys. Res. Lett.\/} {\bf 42}~(11), 4301--4308.

\bibitem[Zhang \& Libchaber(2000)]{zhang2000periodic}
{\sc Zhang, J. \& Libchaber, A.} 2000 Periodic boundary motion in thermal
  turbulence. {\em Physical Review Letters\/} {\bf 84}~(19), 4361.

\bibitem[Zhong \& Zhang(2005)]{Zhong2005}
{\sc Zhong, J.-Q. \& Zhang, J.} 2005 Thermal convection with a freely moving
  top boundary. {\em Phys. Fluids\/} {\bf 17}~(11), 115105.

\bibitem[Zhong \& Zhang(2007{\natexlab{{\em a\/}}})]{zhong2007a}
{\sc Zhong, J.-Q. \& Zhang, J.} 2007{\natexlab{{\em a\/}}} Dynamical states of
  a mobile heat blanket on a thermally convecting fluid. {\em Physical Review
  E\/} {\bf 75}~(5), 055301.

\bibitem[Zhong \& Zhang(2007{\natexlab{{\em b\/}}})]{zhong2007b}
{\sc Zhong, J.-Q. \& Zhang, J.} 2007{\natexlab{{\em b\/}}} Modeling the
  dynamics of a free boundary on turbulent thermal convection. {\em Physical
  Review E\/} {\bf 76}~(1), 016307.

\bibitem[Zhong \& Gurnis(1993)]{zhong1993dynamic}
{\sc Zhong, S. \& Gurnis, M.} 1993 Dynamic feedback between a continentlike
  raft and thermal convection. {\em Journal of Geophysical Research: Solid
  Earth\/} {\bf 98}~(B7), 12219--12232.

\bibitem[Zhong {\em et~al.\/}(2000)Zhong, Zuber, Moresi \&
  Gurnis]{zhong2000role}
{\sc Zhong, S., Zuber, M.~T., Moresi, L. \& Gurnis, M.} 2000 Role of
  temperature-dependent viscosity and surface plates in spherical shell models
  of mantle convection. {\em Journal of Geophysical Research: Solid Earth\/}
  {\bf 105}~(B5), 11063--11082.

\end{thebibliography}

\end{document}